\documentclass[a4paper,11pt]{article}
\usepackage{pos}

\title{Design of a Robust Fiber Optic Communications System for Future IceCube Detectors}
 \ShortTitle{Fiber Test System}

\author{The IceCube Collaboration \\{\normalsize \normalfont(a complete list of authors can be found at the end of the proceedings)\\}}

\emailAdd{rhalliday@icecube.wisc.edu}

\abstract{In this work we discuss ongoing development of a hybrid fiber/copper data and timing infrastructure for the future IceCube-Gen2 detector. The IceCube Neutrino Observatory is a kilometer-scale detector operating with 86 strings of modules. These modules communicate utilizing a custom protocol to mitigate the signaling challenges of long distance copper cables. Moving past the limitations of a copper-based backbone will allow larger future IceCube detectors with extremely precise timing and a large margin of excess throughput to accommodate innovative future modules. To this end, the upcoming IceCube Upgrade offers an opportunity to deploy a pathfinder for the new fiber optic infrastructure, called the Fiber Test System. This design draws on experience from AMANDA and IceCube and incorporates recently matured technologies such as ruggedized fibers and White Rabbit timing to deliver robust and high-performance data and timing transfer.

\vspace{4mm}
{\bfseries Corresponding authors:}
Robert Halliday$^{1*}$, Tyce DeYoung$^{1}$, Chris Ng$^{1}$, Darren Grant$^{1}$, Brian Ferguson$^{1}$, Dean Shooltz$^{1}$\\
{$^{1}$ \itshape Michigan State University, MI 48824, USA}\\[4mm]
$^*$ Presenter

\FullConference{37$^{\rm{th}}$ International Cosmic Ray Conference (ICRC 2021)\\
		July 12th -- 23rd, 2021\\
		Online -- Berlin, Germany}

}

\begin{document}
\maketitle

\section{Background}
The IceCube Neutrino Observatory is a gigaton-scale neutrino detector located at the geographic South Pole. Using more than 5000 photomultiplier tubes housed in Digital Optical Modules (DOMs), IceCube is able to record the Cherenkov light from charged secondary particles originating in interactions with high-energy atmospheric and astrophysical neutrinos. This technique was successful in observing a flux of astrophysical neutrinos, making a more precise measurement of neutrino oscillation parameters and in the first multi-messenger detection of neutrinos from the flaring blazar TXS 0506+056 \cite{hese,joshosc,txs}. Looking forward, the IceCube Upgrade will begin deployment in the 2022/23 austral season, followed by the possible future IceCube-Gen2 detector \cite{upgrade,gen2}. The IceCube Upgrade will provide a powerful low-energy neutrino physics program, a calibration program to improve our understanding of IceCube's systematic uncertainties and a test platform for future devices and technologies for IceCube-Gen2. The IceCube-Gen2 detector is focused on detection of high-energy astrophysical neutrinos and will expand the instrumented volume of IceCube to approximately 8\,km$^3$.

In the course of detecting a neutrino, the IceCube DOMs digitize single photon level or higher signals and send their locally time-tagged information to the surface. The communications for this exchange are done over high quality copper signaling pairs using a custom Amplitude Shift Keying (ASK) protocol for communications, with two DOMs sharing each wire pair. Time tagging and translation is facilitated by the RAPCAL (Reciprocal Active Pulsing Calibration) protocol, which uses a call and response scheme to establish common points in time between a GPS disciplined clock domain in the IceCube Laboratory (on the surface) and each of the in-ice DOMs. The required timing precision in RAPCAL, and in particular the need to ensure quiet on the lines for the duration of the RAPCAL procedure, drives a stringent requirement for cross-talk suppression in IceCube's communication cables. This cross-talk suppression requirement is a major factor in both the engineering and cost of these cables. Through this existing system, IceCube achieves a bandwidth capacity of 720\,kbps per wire pair and a timing accuracy of 1.2\,ns \cite{jinst2017}. In the IceCube Upgrade, DOMs with multiple PMTs will be deployed, raising the requirements on bandwidth capacity per wire pair.

\section{Design Requirements and Rationale for Communications in IceCube-Gen2}
\label{sec:reqs}
In the current IceCube detector, the longest cable runs including surface runs are approximately 3km, while in IceCube-Gen2, the larger detector will require cable runs of up to 6km. At these distances, IceCube's custom ASK protocol will be strained to meet bandwidth, crosstalk and timing requirements, while at the same time the length and purity of the copper cables becomes prohibitively expensive. A natural choice at this point in the design would be to move to fiber optic communications for some or all of that distance. Fiber optic communications at present can support extremely high bandwidths ($>$10\,Gbps) and are deployed in the harshest engineering environments including subsea, petroleum drilling and military aerospace applications. 

Looking back to the AMANDA experiment, a forerunner to IceCube, fiber optics were deployed both for calibration and eventually for signal collection. In the first design deployed, these signals travelled over copper cables of similar length to IceCube, but in the 1999/2000 season, optical modules with loose-tube fibers for signal collection were deployed on 5 strings. After deployment, an Optical Time Domain Reflectometry (OTDR) device was used to probe the integrity of the cables. Measurements indicated that over the course of freeze-in, 5.4\% of fibers failed. Of the fibers that broke, the OTDR data indicate that the breakage was likely at the penetrator interface bringing the fibers into the optical modules \cite{bouchta}. Drawing from this experience, we will require that a future fiber optic communications design both a) uses a more robust fiber and penetrator combination than the AMANDA design and b) that the design continues to operate at full capacity beyond a 5.4\% failure rate of fiber connections.

The ultimate cause of the AMANDA fiber breakages is unclear, and due to the nature of the Antarctic array, will not be directly uncovered. One possible option is that the dynamic pressures of the ice freezing process (referred to as "freeze-in") put uneven pressure on the fibers at the penetrator. In general, IceCube components are rated to 10\,kpsi (69\,MN/m$^2$), which is higher than the observed pressures during freeze-in. We will adopt this pressure rating as an additional requirement for the IceCube-Gen2 Fiber Communications Option.
\begin{figure}[h]
    \centering
    \includegraphics[width= 5 in]{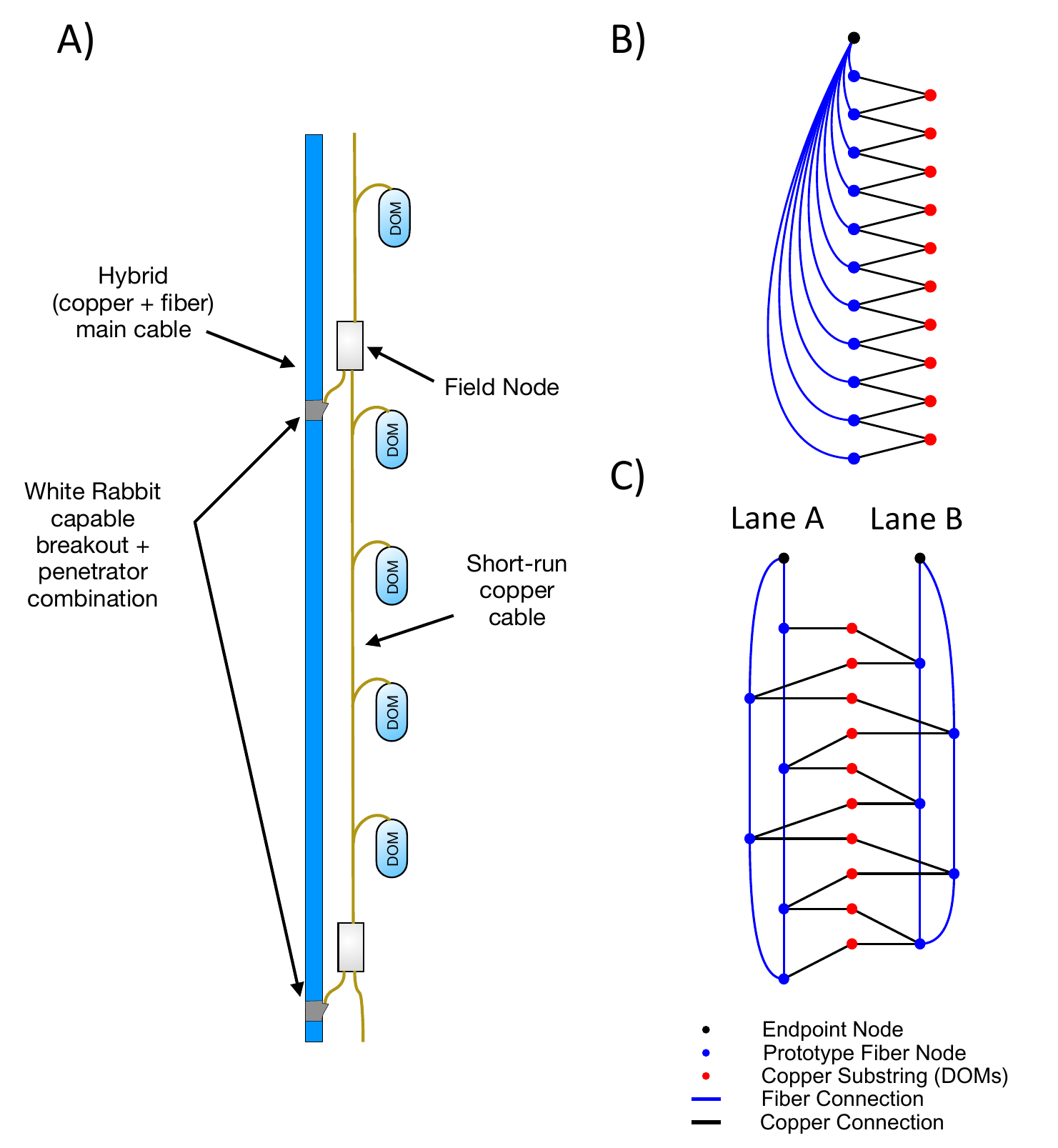}
    \caption{A) Diagram of the overall IceCube-Gen2 Fiber Communications option pointing out main features of the system. B) the "classic" fiber topology, with a single fiber going to each Field Node. C) the interspersed fiber topology, using fewer fibers and achieving greater redundancy than the "classic" topology by dividing connections into two redundant lanes. We also minimize the fiber length between adjacent nodes along each lane, thereby maximizing signal regeneration.}
    \label{fig:mainfig}
\end{figure}
\subsection{Collected Requirements}
\label{subsec:colreqs}
Based on previous experience from AMANDA, and the needs of the IceCube-Gen2 detector, our design for a fiber optic communications system must meet a number of conditions and requirements, namely:
\begin{enumerate}
\item $>$1.5\,Mbps bandwidth capacity per wire pair;
\item $<$1.2\,ns timing accuracy;
\item survives 10\,kpsi pressure testing;
\item full system operation under at least 5\% fiber breakage;
\item more robust fiber and penetrator combination than used in AMANDA; and
\item cost and deployment complexity must be comparable to or less than that of copper cables.
\end{enumerate}

\section{IceCube-Gen2 Fiber Communications Option}

Our proposed fiber communications design, shown in figure \ref{fig:mainfig}, is capable of meeting and exceeding the requirements stated in the preceding sections. The three assemblies containing novel elements in this design are the 1) Fiber Main Cable, 2) the Field Node and 3) the penetrator. The proposed fiber main cable will go without breakouts for the first 1.5-1.7\,km down hole, instrumenting the remaining 1200m down to near the bedrock. In the instrumented region, 11 Field Nodes would be placed every 120\,m along the cable. Field Nodes act as power, data and timing distribution hubs with up to eight sensors redundantly connected to both of the vertically adjacent Field Nodes, ensuring dual routes to the surface. The penetrator is potted into the Field Node and presents a single connection to the main cable. The Field Node will accordingly have two connections for the breakout cables to the Optical Modules. This minimizes the added steps in deployment: breakout cables already are connected every few DOMs, making those connections less frequent and only adding the step of attaching a Field Node in between each breakout cable assembly.  

Physically, the system protects the bulk of the fibers in the main cable, with ruggedized optical transceivers embedded in the breakouts. Inside the length of the main cable, fibers will be armored and bundled with a strength member to handle the vertical tension. In the primary design scheme, the signals converted in the breakout and control signals for the transceivers are sent over a high-conductor-count copper cable with high-speed differential lines to the Field Node for the transceivers' RX and TX signals. This design scheme thereby fully avoids having any fibers go through a pressure-bearing interface at any point, satisfying both the pressure requirement and the requirement to use a more robust system with reference to AMANDA. An alternative design scheme, discussed in subsection \ref{altdesign}, brings the fibers into the Field Node.

To meet the data and timing requirements, we will use the IEEE1588-2019 Annex I.5.3 and Annex M High Precision profile termed "White Rabbit." This protocol delivers timing and bandwidth capacity that easily meet their respective requirements, namely it can deliver pulse-per-second and 10\,MHz signals with jitter on the order of $<$100\,ps and gigabit fiber Ethernet \cite{serrano}. It is widely adopted in the physics community and runs bidirectionally on a single-mode fiber, thereby minimizing the number of fibers needed \cite{wrusers}. Many White Rabbit implementations, including the WR-LEN and CUTE-WR-DP/A7 allow for two connections in support of "cascade mode," inheriting timing from an upstream node and passing it to a single downstream node \cite{len,cute}. As described below, this networking topology can be used to create redundant paths while reducing the number of fibers overall and ensuring even signal regeneration along the path. 

Towards achieving a high degree of redundancy, we choose an internal fiber topology that creates two looped fiber data lanes, regenerating signals at each Field Node. Additionally, the Field Nodes are connected non-sequentially along the fiber lanes, giving a more even distribution of transmission lengths. In comparison with a "classic" topology, where each Field Node has one fiber connected, this "interspersed" topology makes it such that each Field Node and each Optical Module are redundantly connected. As a result, a failure analysis shows that 3 fiber connections, or 23\% of connections can be lost before any sensors fully lose connectivity in the interspersed topology, while the classic design can only accommodate 1 fiber breakage before it starts losing optical modules (depending on the failure configuration). The classic and interspersed topologies are shown in Figure \ref{fig:mainfig}.

Finally, the short-run copper connection will be based on the distributed timing and data acquisition in the  CHIPS and microDAQ systems, which deliver PPS, 10\,MHz, data and power connections over standard Ethernet cables from a fiber optically connected White Rabbit node \cite{chips,microdaq}. Both of these systems are deployed or slated for deployment in similar neutrino and astro-particle detection applications, albeit not in an under-pressure environment. We leave the option of which protocol will be used for data transfer open to be decided at a later time, but either of high speed serial or single pair Ethernet can satisfy per-DOM bandwidth capacity requirements.

\subsection{Penetrator Design Options}
\label{altdesign}
\begin{figure}[h]
    \centering
    \includegraphics[width= 5.95 in]{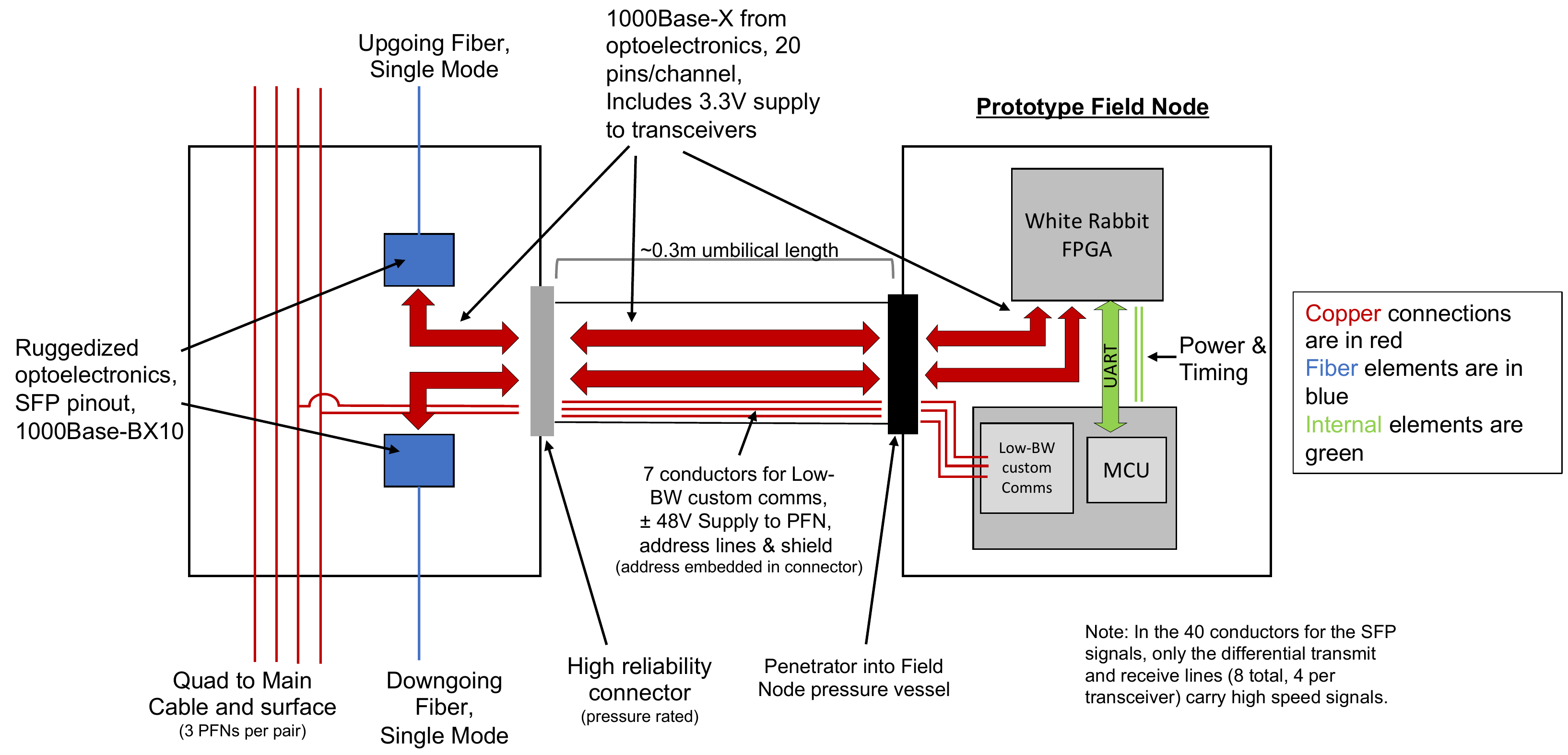}
    \caption{Block diagram of the High Speed Copper breakout and Prototype Field Node design for the Fiber Test System. In contrast with the full-array version shown in Figure \ref{fig:mainfig} panel A, the prototype does not include connections to DOMs but includes a separate controller board for communications with the surface over long-run copper. In the non-prototype Field Node, microcontroller (MCU) functions will be handled on the Zynq chip also housing the White Rabbit functionality.}
    \label{fts_bd}
\end{figure}

As mentioned in the previous section, there are two feasible design options being explored for the penetrator, which brings signals from the Fiber Main Cable to the Field Node. In the first, which we refer to as High Speed Copper (HSC), optoelectronics are embedded in the main cable and converted from 1000Base-BX optical Ethernet signals to 1000Base-X serial electrical Ethernet signals and sent over high speed copper differential lines into the Field Node. A block diagram for a breakout section of a prototype without the local copper connection to DOMs is shown in Figure \ref{fts_bd}. This design keeps all fibers in the Fiber Main Cable where, based on previous experience (see sec. \ref{sec:reqs}), we are confident they will be safe during freeze-in. Alternatively, fibers can be brought into the Field Node in the Hybrid Fiber Penetrator (HFP) design. This design avoids the complications of optoelectronic conversion in the breakout, namely that the breakout will have to be larger diameter than the cable, making spooling it for shipment significantly more difficult. In the HFP design, the breakout is simple, but fibers will ultimately have to go through a pressure-rated interface, and will have to be connected in the field during deployment. A final design option would be to bring the White Rabbit electronics into the breakout. This design was deemed infeasible early on, but in light of recent estimates from vendors on the size of the breakout, we may ultimately pursue this further at a later time.
\begin{figure}[h]
    \centering
    \includegraphics[width= 5.95 in]{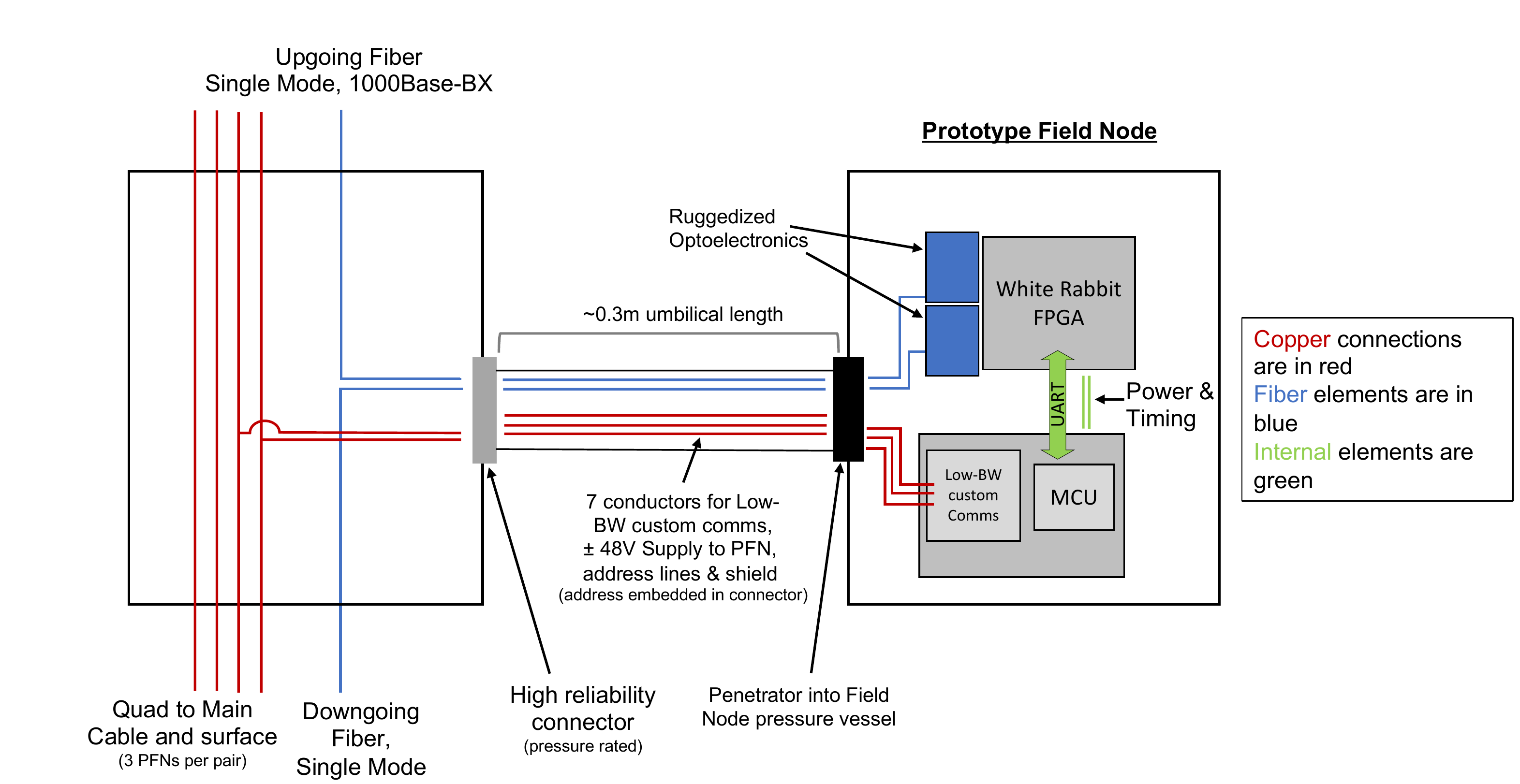}
    \caption{Block diagram of the Hybrid Fiber Penetrator breakout and Prototype Field Node design for the Fiber Test System. In this design, fibers running to the Prototype Field Node simplify optoelectronic conversion but add the risk of connector flooding, and the complexity of making fiber optic connections in South Pole conditions.}
    \label{hfp}
\end{figure}
\section{Fiber Test System}
\label{sec:fts}
In the upcoming IceCube Upgrade, slotted for deployment in the 2022/23 austral summer, there is an opportunity to deploy a pathfinder system for the IceCube-Gen2 Fiber Communications option. This system, dubbed the "Fiber Test System," will consist of six Prototype Field Nodes, attached to a Fiber Test Cable through one of the two proposed penetrator choices (HSC or HFP). This system will function as a drop-in replacement for an IceCube Upgrade DOM breakout cable, corresponding to 4 DOMs. 

The Fiber Test Cable will carry standard IceCube Upgrade copper cables for communications, power and timing with the surface, and an additional single fiber lane so the Prototype Field Nodes can communicate amongst themselves. At this point, we are primarily pursuing the HSC design, so optoelectronic conversion will occur in the breakout with ruggedized transceivers. We see the HSC design as marginally more safe since it avoids sending fibers through pressure rated connectors, thereby avoiding the possibility of flooding or difficult deployment conditions causing optical misalignment. A breakout and Prototype Field Node for this design is shown in Figure \ref{fts_bd} as a block diagram.

Testing is a particularly important part of this pathfinder program. All parts will be fully tested to 10\,kpsi before acceptance, and tested for electrical and optical continuity and electrical isolation before shipping and again before deployment. In order to gain the most data from this deployment, the system's integrity should be known at each point in the delivery chain.

\subsection{Field Node electronics and White Rabbit Design}
The White Rabbit Node to be deployed in the Fiber Test System is targeted to be a two SFP instantiation of the White Rabbit PTP Core on a Xilinx Zynq SoC/FPGA or a Xilinx Kintex FPGA. We are targeting the Avnet PicoZed 7030 SOM in the industrial grade, which should cover the operating temperature in the South Pole ice. This PicoZed would be mounted on a daughterboard with cages for SFP style connections, White Rabbit clocking resources, and a ribbon header to be connected to the computing bundle's MCU. In light of the global shortage of PicoZed boards, we are exploring design alternatives including a board based on the CUTE-WR-A7.

Traffic handling and slow control will occur on the Zynq's dual-core ARM processor, while timing distribution will be sourced from the White Rabbit PTP Core. A design for power distribution to the sensors has not been chosen; we are reviewing options for fail-safe microcontrollers which would allow parallel power distribution off of a minimum number of heavy conductors, or the use of lower quality conductors for individual power distribution to each Field Node. In either configuration, the resulting cable will have fewer conductors and should therefore have a smaller diameter than current IceCube cables.  

\section{Conclusions and Outlook}
In this work, we have compiled a selection of requirements and considerations for a future fiber optic communications system for IceCube-Gen2 or any large-scale in-ice or subsea detector array. Engineering of the mechanical aspects of the design and optoelectronic conversion in the breakout satisfy concerns about fiber exposure and integrity under pressure, while the application of the White Rabbit protocol meets and exceeds the timing and data throughput requirements for IceCube-Gen2. By appropriately arranging fibers within the communications system, we can ensure redundancy, allowing the breakage of, at minimum, 23\% of fiber connections without losing communications to a single sensor module. The culmination of these developments enables a large scale fiber optic communication design that bypasses the difficulties of copper signaling, provides wide redundancies and meets or exceeds all data and timing requirements, thereby paving the way for more advanced detectors and further reaching scientific goals. Towards testing the physical and electrical components of such a system, we will deploy the Fiber Test System in the IceCube Upgrade, which will give us critical validation data for the implementation of a next-generation fiber communications and timing design.

\bibliographystyle{ICRC}
\bibliography{references}



\clearpage
\section*{Full Author List: IceCube Collaboration}




\scriptsize
\noindent
R. Abbasi$^{17}$,
M. Ackermann$^{59}$,
J. Adams$^{18}$,
J. A. Aguilar$^{12}$,
M. Ahlers$^{22}$,
M. Ahrens$^{50}$,
C. Alispach$^{28}$,
A. A. Alves Jr.$^{31}$,
N. M. Amin$^{42}$,
R. An$^{14}$
% ...
\\

\noindent
% ...
$^{12}$ Universit{\'e} Libre de Bruxelles, Science Faculty CP230, B-1050 Brussels, Belgium \\
% ...
$^{14}$ Department of Physics and Laboratory for Particle Physics and Cosmology, Harvard University, Cambridge, MA 02138, USA \\
% ...
$^{17}$ Department of Physics, Loyola University Chicago, Chicago, IL 60660, USA \\
$^{18}$ Dept. of Physics and Astronomy, University of Canterbury, Private Bag 4800, Christchurch, New Zealand \\
% ...
$^{22}$ Niels Bohr Institute, University of Copenhagen, DK-2100 Copenhagen, Denmark \\
% ...
$^{28}$ D{\'e}partement de physique nucl{\'e}aire et corpusculaire, Universit{\'e} de Gen{\`e}ve, CH-1211 Gen{\`e}ve, Switzerland \\
% ...
$^{31}$ Karlsruhe Institute of Technology, Institute for Astroparticle Physics, D-76021 Karlsruhe, Germany  \\
% ...
$^{42}$ Bartol Research Institute and Dept. of Physics and Astronomy, University of Delaware, Newark, DE 19716, USA \\
% ...
$^{50}$ Oskar Klein Centre and Dept. of Physics, Stockholm University, SE-10691 Stockholm, Sweden \\
% ...
$^{59}$ DESY, D-15738 Zeuthen, Germany \\
% ...

\subsection*{Acknowledgements}

\noindent
USA {\textendash} U.S. National Science Foundation-Office of Polar Programs,
U.S. National Science Foundation-Physics Division,
U.S. National Science Foundation-EPSCoR,
Wisconsin Alumni Research Foundation,
Center for High Throughput Computing (CHTC) at the University of Wisconsin{\textendash}Madison,
Open Science Grid (OSG),
Extreme Science and Engineering Discovery Environment (XSEDE),
Frontera computing project at the Texas Advanced Computing Center,
U.S. Department of Energy-National Energy Research Scientific Computing Center,
Particle astrophysics research computing center at the University of Maryland,
Institute for Cyber-Enabled Research at Michigan State University,
and Astroparticle physics computational facility at Marquette University;
Belgium {\textendash} Funds for Scientific Research (FRS-FNRS and FWO),
FWO Odysseus and Big Science programmes,
and Belgian Federal Science Policy Office (Belspo);
Germany {\textendash} Bundesministerium f{\"u}r Bildung und Forschung (BMBF),
Deutsche Forschungsgemeinschaft (DFG),
Helmholtz Alliance for Astroparticle Physics (HAP),
Initiative and Networking Fund of the Helmholtz Association,
Deutsches Elektronen Synchrotron (DESY),
and High Performance Computing cluster of the RWTH Aachen;
Sweden {\textendash} Swedish Research Council,
Swedish Polar Research Secretariat,
Swedish National Infrastructure for Computing (SNIC),
and Knut and Alice Wallenberg Foundation;
Australia {\textendash} Australian Research Council;
Canada {\textendash} Natural Sciences and Engineering Research Council of Canada,
Calcul Qu{\'e}bec, Compute Ontario, Canada Foundation for Innovation, WestGrid, and Compute Canada;
Denmark {\textendash} Villum Fonden and Carlsberg Foundation;
New Zealand {\textendash} Marsden Fund;
Japan {\textendash} Japan Society for Promotion of Science (JSPS)
and Institute for Global Prominent Research (IGPR) of Chiba University;
Korea {\textendash} National Research Foundation of Korea (NRF);
Switzerland {\textendash} Swiss National Science Foundation (SNSF);
United Kingdom {\textendash} Department of Physics, University of Oxford.




\end{document}